\documentstyle[prb,aps]{revtex}
\begin{document}
\draft
\twocolumn[\hsize\textwidth\columnwidth\hsize\csname
@twocolumnfalse\endcsname
\title{ Hole-hole superconducting pairing in the $t$-$J$ model
         induced by spin-wave exchange}
\author{V. I. Belinicher and A. L. Chernyshev}
\address{Institute of Semiconductor Physics, 630090 Novosibirsk, Russia}

\author{A. V. Dotsenko}
\address{School of Physics, The University of New South Wales, Sydney 2052,
    Australia}
\author{O. P. Sushkov}
\address{School of Physics, The University of New South Wales, Sydney 2052,
    Australia\\
  and Budker Institute of Nuclear Physics, 630090 Novosibirsk, Russia}

\date{{\tt cond-mat/9406039}; to be published in Phys.\ Rev.\ B}

\maketitle
\begin{abstract}

We study numerically the hole pairing induced by spin-wave exchange.
The contact hole-hole interaction is taken into account as well.
It is assumed that antiferromagnetic order is preserved
 at all scales relevant to pairing.
The strongest pairing is obtained for the $d$-wave symmetry of the gap.
Dependence of the value of the gap on hole concentration and temperature
 is presented.
For the critical temperature we obtain
 $T_c \sim 100$  K at the hole concentration  $\delta \sim $0.2--0.3.
\end{abstract}

\pacs{
  74.20.Mn, 
  74.20.Fg, 
  71.27.+a, 
  74.25.Jb} 
\vskip2pc]

\section{Introduction}
\noindent
Magnetic fluctuations are believed to be a very likely
  mechanism of pairing in cuprate superconductors.
There have been many
 studies\cite{Sch9,Bic9,Kam1,Kam2,Bul1,Bul2,Mon1,Mon2,Dah3}
 of predominantly phenomenological nature  supporting this idea.
In the present work we study spin-wave-mediated hole pairing
  using results obtained from {\em first principles} for
 the undoped $t$-$J$ model.

We base our study  on the results of previous papers.\cite{Kuch3,Fla4}
It was shown in Ref.\onlinecite{Kuch3} that because of spin-wave
  exchange there is an effective long-range attraction between two
  holes with opposite spins,
\begin{equation} \label{pot}
   U_{\rm eff}(r) = \frac{\lambda}{r^2},\hspace{1.cm}\lambda < 0.
\end{equation}
In this potential there is an infinite series of two-hole bound states.
However, they have very large sizes and very small binding energies and
 thus are not directly responsible for high-$T_c$ superconductivity.
Very strong pairing in the many-hole problem due to the
 same potential (\ref{pot}) was demonstrated in Ref. \onlinecite{Fla4},
 where an infinite set of solutions for the superconducting gap was found.
The strongest pairing was either in the $d$-wave or $g$-wave sector.
The pairing induced by spin-wave exchange is a long-range phenomenon.
However, the attractive potential (\ref{pot}) is too
 singular and the wave function is known to collapse to the origin.
On the one hand, this ``collapse'' effect substantially enhances pairing.
On the other hand, it leads to a dependence of the superconducting gap
 on short-range dynamics which cannot be studied analytically.
For this reason analytical calculations\cite{Fla4} can only estimate
 the numerical value of the gap and cannot
 distinguish between $d$- and $g$-wave pairing (which have the same
 long-range behavior but a different short-range one).
In the present work, we calculate the gap numerically, taking into account
 both spin-wave exchange and contact hole-hole interaction.
The $d$-wave pairing is shown to be the strongest.

Recently,  $d$-wave pairing was studied in
 Ref.~\onlinecite{Dag5}.
Although many results are similar,
 we believe the spin-wave exchange interaction which we use
 is more realistic  than the atomic limit interaction employed
 in Ref.~\onlinecite{Dag5}.

Our paper has the following structure.
In Sec. II we present an effective Hamiltonian of the $t$-$J$ model.
In Sec. III we calculate the BCS-type pairing of holes at zero temperature.
Section IV presents the results of the calculation of the critical
 temperature.
Finally, our conclusions are given in Sec.~V.

\section{Effective Hamiltonian for \protect\\ dressed holes}
\noindent
The underlying microscopic physics is described by the $t$-$J$ model
 defined by the Hamiltonian
\begin{eqnarray} \label{H}
  H &=& H_t + H_J \nonumber\\
    &=& -t \sum_{\langle nm\rangle\sigma}
            ( d_{n\sigma}^{\dag} d_{m\sigma} + \mbox{H.c.} )
     + J \sum_{\langle nm\rangle}  {\bf S}_n \cdot {\bf S}_m,
\end{eqnarray}
where $d_{n\sigma}^{\dag}$ is the creation operator of a hole with spin
 $\sigma$ ($\sigma= \uparrow, \downarrow$) at site $n$ on a
 two-dimensional square lattice.
The $d_{n\sigma}^{\dag}$ operators act in Hilbert space
 with no double electron occupancy.
The spin operator is ${\bf S}_n = {1 \over 2} d_{n \alpha}^{\dag}
 ${\boldmath $\sigma$}$_{\alpha \beta} d_{n \beta}$.
$\langle nm\rangle$ are the nearest-neighbor sites on the lattice.
Below we set $J=1$ and give all energy values in units of $J$.

At half-filling (one hole per site) the $t$-$J$ model is equivalent to
 the Heisenberg antiferromagnet model which has long-range N\'{e}el
 order in the ground state.
Under doping the long-range antiferromagnetic order is destroyed.
However, local antiferromagnetic order is preserved.
We assume that the magnetic correlation length $\xi_{\rm magn}$
 is not smaller than  the typical wavelength of holes,
 $\xi_{\rm magn} \ge 1/p_F \sim 1/\sqrt{\delta}$
 ($\delta \ll 1$ is the concentration of holes).
Thus we have antiferromagnetic order at all scales relevant to the problem.
This assumption does not contradict experimental data.\cite{Birg}

We treat the $J$ term of the Hamiltonian (\ref{H}) using the linear spin-
  wave approximation  (see Ref.~\onlinecite{Manousakis} for a review).
Define the Fourier transformations
\begin{equation} \label{2}
 a_{\bf q}^{\dag} =  \sqrt{2\over{N}} \sum_{n \in \uparrow}
                 S_n^- e^{i{\bf q} \cdot {\bf r}_n},
\hspace{0.5cm}
  b_{\bf q}^{\dag} = \sqrt{2\over{N}} \sum_{n \in \downarrow}
                 S_n^+ e^{i{\bf q} \cdot {\bf r}_n},
\end{equation}
where the notation  $n \in \uparrow$ ($n \in \downarrow$) means that
 site $n$ is on the spin-up (-down) sublattice.
Introducing the Bogolubov canonical transformation
\begin{equation} \label{3}
\alpha_{\bf q}^{\dag} = U_{\bf q} a_{\bf q}^{\dag} -
                        V_{\bf q} b_{\bf -q}, \hspace{0.5cm}
\beta_{\bf q}^{\dag} = U_{\bf q} b_{\bf q}^{\dag} -
                       V_{\bf q} a_{\bf -q},
\end{equation}
we write the Heisenberg Hamiltonian $H_J$ as
\begin{equation} \label{4}
  H_J = E_0 + \sum_{\bf q}  \omega_{\bf q}
  (\alpha_{\bf q}^{\dag}\alpha_{\bf q} + \beta_{\bf q}^{\dag}\beta_{\bf q}),
\end{equation}
 where $E_0$ is the antiferromagnetic background energy.
The summation over ${\bf q}$ is restricted to the Brillouin
 zone of one sublattice
 [$\gamma_{\bf q} = {1\over 2} (\cos q_x+\cos q_y)\ge 0$].
The spin-wave dispersion and the transformation coefficients are given by
\begin{eqnarray} \label{5}
  \omega_{\bf q} = 2\sqrt{1-\gamma_{\bf q}^2}, &\hspace{0.5cm}&
  \omega_{\bf q} \approx  \sqrt{2}|{\bf q}|
  \ \mbox{at} \ |{\bf q}|  \ll 1,   \nonumber \\
     U_{\bf q} = \sqrt{{1\over{\omega_{\bf q}}}+{1\over{2}}},
 &\hspace{0.5cm}&
     V_{\bf q} = -\mbox{sgn}(\gamma_{\bf q})
                   \sqrt{{1\over{\omega_{\bf q}}}-{1\over{2}}}.
\end{eqnarray}
The spin waves created
  by $\alpha_{\bf q}^{\dag}$ and $\beta_{\bf q}^{\dag}$
  have definite values of spin projection.
Due to Eqs. (\ref{2}) and (\ref{3}),
 $\alpha_{\bf q}^{\dag} |0\rangle$ has $S_z=-1$ and
  $\beta_{\bf q}^{\dag} |0\rangle$ has $S_z=+1$.
Here $|0\rangle$ is the wave function of the quantum N\'{e}el state.

Single-particle properties in the $t$-$J$ model are by now well
 established (see Ref.~\onlinecite{Dag4}).
A single hole is
 a magnetic polaron of a small radius, i.e., a ``bare''
 hole that is ``dressed'' by virtual spin excitations.
A single hole has a ground state
 with a momentum of ${\bf k}=(\pm \pi/2, \pm\pi/2)$.
The energy is almost degenerate along the line $\cos k_x+\cos k_y=0$
 which is the edge of the magnetic Brillouin zone
 (see, e.g., Refs. \onlinecite{Kane,Dag0,Mart1,Liu2,Sus1,Sus2,She2,Giam3}).
The hole dispersion may be well approximated by the
 analytical expression\cite{Sus2}
\begin{eqnarray} \label{6}
 \epsilon_{\bf k} &=& \sqrt{\Delta_0^2/4+4t^2(1+y)} \nonumber\\
  &-& \sqrt{\Delta_0^2/4+4t^2(1+y)-4t^2(x+y)\gamma_{\bf k}^2} \nonumber\\
  &+& {1\over4} \beta_2 (\cos k_x -\cos k_y)^2; \nonumber \\
  \Delta_0 &\approx& 1.33,\ x \approx 0.56, \ y \approx 0.14,
\end{eqnarray}
where the parameters $\Delta_0,x,y$ are some combinations of the ground
 state spin correlators.\cite{Sus2}
Near the band bottom ${\bf k}_0=(\pm \pi/2, \pm \pi/2)$
 the dispersion (\ref{6}) can be presented in the usual quadratic form
\begin{equation} \label{7}
\epsilon_{\bf p} \approx {1\over2} \beta_1 p_1^2 + {1\over2} \beta_2 p_2^2,
 \hspace{1.0cm} \beta_2 \ll \beta_1,
\end{equation}
where $p_1$ ($p_2$) is the projection of ${\bf k}-{\bf k}_0$
 on the direction orthogonal (parallel) to the face of the
 magnetic Brillouin zone (Fig.~1).
{}From Eq.~(\ref{6}) for $t \gg \Delta_0 /4$ we have
\begin{equation} \label{beta1}
 \beta_1 = { {x+y} \over {\sqrt{1+y}} }t \approx 0.65 t.
\end{equation}
According to Refs. \onlinecite{Mart1} and \onlinecite{Giam3},
 $\beta_2 \approx 0.1 t$ at $t \ge \Delta_0/4$.
The wave function of a single hole may be written in the form
 $\psi_{{\bf k}\sigma} = h_{{\bf k}\sigma}^{\dag} |0\rangle$.
At large $t$ the composite hole operator $h_{{\bf k}\sigma}^{\dag}$ has
 complex structure.
For example, at $t/J=3$, very roughly, the weight of
 a bare hole in $\psi_{{\bf k} \sigma}$
 is  about 25\%; the weight of configurations ``bare hole + 1 magnon''
  is  $\sim$50\%, and of configurations ``bare hole + 2 or more
  magnons'' $\sim$25\%.
These estimations are based on the approach
 with a minimal string ansatz \cite{Sus2}
 and further renormalization due to additional magnons.\cite{Suhf}
However, other approaches like finite cluster diagonalization
 \cite{Dag0} or numerical solution of Dyson's equation
 \cite{Kane,Mart1,Liu2,She2} give very close results.
We have to stress that the dressed hole is a normal fermion.

The interaction of a composite hole with  spin waves
 is of the form (see, e.g., Refs.\onlinecite{Mart1,Liu2,Suh3,Suhf})
\begin{eqnarray} \label{11}
H_{\rm h,sw} &=& \sum_{\bf k,q} g_{\bf k,q}
    \left(
  h_{{\bf k}+{\bf q}\downarrow}^{\dag} h_{{\bf k}\uparrow} \alpha_{\bf q}
  + h_{{\bf k}+{\bf q}\uparrow}^{\dag} h_{{\bf k}\downarrow} \beta_{\bf q}
  + \mbox{H.c.}   \right),\nonumber \\
 g_{\bf k,q} &=& 2f \ \sqrt{2\over N}
 (\gamma_{\bf k} U_{\bf q} + \gamma_{{\bf k}+{\bf q}} V_{\bf q}).
\end{eqnarray}
For arbitrary $t$ the coupling constant $f$
 was calculated in Refs.\onlinecite{Suh3,Suhf}.
The plot of $f$ as a function of $t$ is presented in Fig.~2.
For large $t$ the coupling constant is $t$-independent $f \approx 2$.

Let us stress that even for $t > J$ the interaction (\ref{11}) between
   quasiholes and spin waves has the form as
   for $t \ll J$ (i.~e., as for bare hole  operators)
   with an added renormalization factor
   (of the order of $J/t$ for $t \gg J$).
This is a remarkable property of the $t$-$J$ model which is
 due to the absence
 of a single-loop correction to the vertex.
This property was first found perhaps in Ref.~\onlinecite{Kane}.
In Refs.~\onlinecite{Mart1,Liu2,Suhf} it was
 demonstrated explicitly that vertex corrections with different kinematic
 structure are of the order of few percent at $t/J \approx 3$.
There is also a weak $q$-dependence of the coupling constant $f$.
The plot in Fig.~2 corresponds to the long-wavelength limit $q=0$
 because, as we will see later,
 the small $q$'s are most important for pairing.
At $q\sim\pi$, the factor $f$ is $(10-17)$ \% bigger
  than at $q=0$
 [see the discussion between Eqs.\ (13) and (14)
  of Ref.~\onlinecite{Kuch3}].
The influence of this correction on the pairing is negligible.

\onecolumn
Note that the hole scattering
  between different pockets makes a large contribution to the pairing.
However, in the two-sublattice formalism which we use, there are no
  spin waves with $ {\bf q} = {\bf g} = (\pm \pi, \pm \pi) $ and such
  scattering takes place via umklapp processes with $q \sim p_F \ll 1$.
One could use another description:
Expand the Brillouin
 zone for spin waves and include ${\bf q}\approx {\bf g}$ into
 consideration explicitly.
Then, due to antiferromagnetic order
 the points ${\bf q}=0$ and ${\bf q}={\bf g}$ are equivalent, and
 the coupling constants in the effective Hamiltonian (\ref{11})
 are exactly equal, $f_{{\bf q}=0} = f_{{\bf q}={\bf g}}$.
Certainly the kinematic structure of the vertex
 (\ref{11}) reflects this symmetry: $g_{\bf k,q}=g_{\bf k,q+g}$.

Interaction between two holes can be caused by exchange of one spin wave.
Alongside that there is a contact hole-hole interaction.
One can say that it is due to exchange of several
  hard spin-wave excitations.
The Hamiltonian of the contact hole-hole interaction was derived in
 Refs.~\onlinecite{Cher3,Kuch3} using a variational approach:
\begin{eqnarray} \label{14}
H_{\rm h,h} &=& {8\over N} \sum_{{\bf k}_1,{\bf k}_2,{\bf k}_3,{\bf k}_4}
 \biggl[
  A\gamma_{{\bf k_1}-{\bf k_3}}
 + {1\over 2}C (\gamma_{{\bf k}_1+{\bf k}_3}+\gamma_{{\bf k}_2+{\bf k}_4})
    \biggr]
    h_{{\bf k}_3 \uparrow}^{\dag} h_{{\bf k}_4 \downarrow}^{\dag}
    h_{{\bf k}_2 \downarrow} h_{{\bf k}_1 \uparrow}
   \delta_{ {\bf k}_1+{\bf k}_2 , {\bf k}_3+{\bf k}_4 },\\
A &=& 16t\nu \mu^3(1-7\mu^2) - {1\over4}-2\mu^2-18.5\mu^4+84\mu^6 +
 10\alpha t \nu^3 \mu^3, \hspace{0.5cm}
     C = {2\over 3}\alpha t \nu \mu^3,\nonumber
\end{eqnarray}
where
\begin{equation} \label{10}
 \nu = {1\over 2} \biggl[ {{3/2+2S_t}\over S_t} \biggr]^{1/2},
    \hspace{.7cm}
 \mu = {t \over {[S_t(3/2+2S_t)]^{1/2}} }, \hspace{.7cm}
 S_t = [9/16+4t^2]^{1/2}.
\end{equation}
The coefficients $A$ and $C$ in Eq.~(\ref{14}) were derived in
  first order in $\alpha$, where $\alpha$ is the coefficient in front of
 the transverse part to the Heisenberg interaction:
 ${\bf S}_n{\bf S}_m \to S^z_n  S^z_m +
 {\alpha \over 2} ( S^+_n S^-_m + S^-_n S^+_m )$.
Since the physical value is $\alpha=1$,
 contributions of higher orders are important.
In order to estimate them, we will set $\alpha=0.6$.
This choice is made so that results for the binding energy of
 short-range two-hole bound states obtained by finite lattice
 diagonalizations\cite{Ede2,Bon2,Poilblanc,PRD3,Sher}
 would agree with results obtained\cite{Cher3,Kuch3}
 by using the effective interaction (\ref{14}).
Actually at $\alpha=0.6$
 the contact interaction $H_{\rm h,h}$ is very small and practically
 does not influence the pairing.

To summarize, we conclude that the dynamics of holes on
 the antiferromagnetic background is described by the effective Hamiltonian
\begin{equation} \label{15}
H_{\rm eff} =
    \sum_{{\bf k} \sigma}
     \epsilon_{\bf k} h_{{\bf k}\sigma}^{\dag} h_{{\bf k}\sigma}
   + \sum_{\bf q}
     \omega_{\bf q} (\alpha_{\bf q}^{\dag} \alpha_{\bf q}
                     +\beta_{\bf q}^{\dag} \beta_{\bf q})
  + H_{\rm h,sw} + H_{\rm h,h},
\end{equation}
which is expressed in terms of the composite hole $h_{{\bf k}\sigma}$
  and spin-wave $\alpha_{{\bf q}}, \beta_{{\bf q}}$ operators.
It includes free holes and spin waves and their interactions
 $H_{\rm h,sw}$ and $H_{\rm h,h}$
 [given by Eqs. (\ref{11}) and (\ref{14})].

\section{Superconducting state}
\noindent
For the small concentrations $\delta \ll 1$ under consideration,
 holes are localized in momentum space in
 the vicinity of the minima of the
 band, ${\bf k}_0 = (\pm \pi/2, \pm \pi/2),$ and
 the Fermi surface consists of ellipses (see Fig.~1).
The Fermi energy and Fermi momentum of noninteracting holes are
\begin{equation} \label{ep}
 \epsilon_F= \frac{1}{2} \pi (\beta_1 \beta_2)^{1/2} \delta, \qquad p_F
    \sim (\pi \delta)^{1/2}.
\end{equation}
The Fermi momentum $p_F$ is measured
  from the center of the corresponding ellipse.
Let us stress that the numerical value of $\epsilon_F$ is very small.
For realistic superconductors $t/J \approx 3$
 (see, e.g., Refs.~\onlinecite{Esk0,Fla1,BCh3}).
Therefore at $\delta=0.1$ and $J=0.15$ eV one gets
 $\epsilon_F\approx 15$ meV $\approx 175$ K.
In pairing, the exchange of spin waves with typical
  momentum $q \sim p_F \ll 1$ is the most important.
The energy of
  such spin waves is much higher than the typical energy of a pair,
\begin{equation} \label{omega}
\omega_q \sim p_F \sim (\pi \delta)^{1/2} \gg \epsilon_F \sim (\beta_1
\beta_2)^{1/2} \delta.
\end{equation}
The situation is quite similar to that for the two-hole bound state
 problem\cite{Kuch3} and much different from the situation with the usual
 phonon-induced pairing where Debye's frequency is much lower than
 the Fermi energy.

The interaction between two holes with opposite spins and
 opposite momenta is\cite{Kuch3}
\begin{equation}
\label{Vkk}
V_{\bf k, k'} = -2 {
      { g_{\bf k, q} g_{\bf k', -q} }
   \over
     { -\omega_{\bf q} - E_{\bf k} - E_{\bf k'} }
   }
+ {8\over N} ( A\gamma_{\bf k-k'} + C\gamma_{\bf k+k'} ).
\end{equation}
\twocolumn
The first term here is due to the spin-wave exchange diagrams
 shown in Fig.~3.
The minus sign before this term takes into account the fact that
 spin-wave exchange makes the spin flip for both holes.
For the same reason, the momentum transfer is the sum (not the difference)
 of the hole momenta ${\bf q}= {\bf k}+ {\bf k}'$.
The energy denominator in Eq.~(\ref{Vkk}) takes into
 account the energy of the spin-wave $\omega_{\bf q}$, and the energies
 $E_{\bf k}$ and $E_{\bf k'}$
 of the two holes in intermediate unpaired state.
In fact, the account of $E_{\bf k}$ and $E_{\bf k'}$ is the account of
 retardation.
We discuss this question below.
The second term in Eq.~(\ref{Vkk}) is the contact interaction (\ref{14}).

We use the usual BCS wave function for the ground state of the many-hole
  system
\begin{equation} \label{Psi}
 | \Psi \rangle = \prod_{\bf k} (u_{\bf k}+v_{\bf k}
   h^{\dag}_{{\bf k}\uparrow} h^{\dag}_{-{\bf k}\downarrow} ) |0 \rangle.
\end{equation}
Thus we suppose that all quasiparticles are in the condensate.
For strong interactions the validity of this assumption is under
 question because there is no parameter to justify it.
We believe that numerically the wave function (\ref{Psi}) is good.
Anyway one may consider the wave function (\ref{Psi}) as a trial one in the
 variational method.
In this case the large gain in energy which we get is a justification
 of the wave function.

The gap $\Delta_{\bf k}$  corresponding to the wave function (\ref{Psi})
 satisfies the conventional BCS equation
\begin{equation} \label{Delta}
 \Delta_{\bf k} = -
  \frac{1}{2} \sum_{\bf k'} V_{\bf kk'} \frac{\Delta_{\bf k'}}
  {\sqrt{\xi_{\bf k'}^2+ \Delta_{\bf k'}^2}},
\end{equation}
where
$\xi_{\bf k}=\epsilon_{\bf k}- \mu$, $\mu$ being the the chemical potential
 fixed by the hole density
\begin{equation} \label{dens}
  \delta = 2 \sum_{\bf k} v_{\bf k}^2.
\end{equation}
It is well known that the excitation energy of fermions in BCS theory is
 $E_{\bf k} = \sqrt{\xi_{\bf k}^2+ \Delta_{\bf k}^2}$.
Just this energy enters Eq.~(\ref{Vkk})
 for the effective hole-hole interaction.
Equation~(\ref{Delta}) is obtained by variation of the average value of the
 Hamiltonian with respect to the parameters $u_{\bf k}$ and $v_{\bf k},$
\begin{equation} \label{var}
  {{\delta}\over{\delta u_{\bf k}}}\langle \Psi|H-{\cal E}|\Psi \rangle=
  {{\delta}\over{\delta v_{\bf k}}}\langle \Psi|H-{\cal E}|\Psi \rangle=0.
\end{equation}
Here ${\cal E}$ is the energy of the ground state.
The effective interaction (\ref{Vkk}) itself depends on the
 parameters $u_{\bf k}$ and $v_{\bf k}$ via the dependence of $E_{\bf k}$
 on the gap $\Delta_{\bf k}$.
Nevertheless, in the variational equations (\ref{var}) we have to set
\begin{equation}\label{Vvar}
   {{\delta}\over{\delta u_{\bf p}}} V_{\bf kk'}=
   {{\delta}\over{\delta v_{\bf p}}} V_{\bf kk'}=0,
\end{equation}
 and therefore we get the usual BCS equation (\ref{Delta}).
Explanation of the condition (\ref{Vvar}) is as follows.
The spin-wave exchange part of the interaction (\ref{Vkk}) is
 due to the second order of perturbation theory.
Therefore, the actual denominator in the spin-wave contribution is
 ${\cal E}-{\cal E}_{\rm excited}$, and it does not depend explicitly on
 $u_{\bf k}$ and $v_{\bf k}$.
The self-consistency condition ${\cal E}-{\cal E}_{\rm excited} =
  - \omega_{\bf q} - E_{\bf k} - E_{\bf k'}$
 appears after solving Eqs.~(\ref{var}) and (\ref{Vvar}).
{}From the practical point of view this question is not important because
  due to the condition (\ref{omega}) the dependence of
  $V_{\bf k,k'}$ on the gap is very weak.

An iterative numerical solution of Eq.~(\ref{Delta}) is straightforward.
We present results for $t=3$ corresponding to realistic
 superconducting systems.\cite{Esk0,Fla1,BCh3}
Since the inverse mass $\beta_2$ [see Eqs.~(\ref{6}),(\ref{7})]
 is known with rather poor accuracy,
 we use several values of the mass ratio $a=\beta_1/\beta_2$.
We take $\beta_1$ from Eq.~(\ref{beta1}) and then set $\beta_2=\beta_1/a$.
The constant of the hole-magnon interaction (\ref{11}) is $f=1.80$ at $t=3$.

The symmetry group of the square lattice is $C_{4v}$.
The solutions of Eq.~(\ref{Delta}) belong to certain representations
  of this group.
In agreement with Ref.~\onlinecite{Fla4}, the strongest pairing
  is in the $B_1$ representation [$d$ wave, Fig.~4(a)]
 and in the $A_2$ representation [$g$ wave, Fig.~4(b)].
Consider first the $d$-wave pairing.
The map of the gap for the hole concentration $\delta=0.1$ and
 the mass ratio $a=\beta_1/\beta_2=7$ is presented in Fig.~5.
In Fig.~6 for the same parameters we give the map of $v_{\bf k}^2$
 which is the mean occupation number of a single-hole quantum state.
We observe that despite a big value of gap the hole density
 distribution changes quite sharply at crossing the Fermi surface.
For other mass ratios and hole concentrations
 Fig.~5 is also approximately valid because it gives
 the gap in units of $\Delta_{\rm max}$ and we found that
 with changing $\delta$ and $a$, the whole gap function is
 multiplied by some factor but the ${\bf k}$ dependence is
 not much changed,
\begin{equation} \label{scaling}
 \Delta_{\bf k} (\delta, a) \approx
 { \Delta_{\rm max} (\delta, a) \over \Delta_{\rm max} (\delta', a') }
 \Delta_{\bf k} (\delta', a').
\end{equation}
Due to interaction between holes, the ideal gas relation (\ref{ep})
 between the chemical potential $\mu$ and
 the hole concentration is not valid.
The correct relation follows from Eq.~(\ref{dens}).
Plots of $\mu$ as a function of $\delta$ are given in Fig.~7.

A very important characteristic is the maximam value
 of the gap on the Fermi surface, $\Delta_1.$
Its value is directly related
 to the critical temperature of the superconducting
 transition,\cite{Fla4}
\begin{equation} \label{dtc}
  T_c \approx 0.5 \Delta_1(T=0).
\end{equation}
According to Fig.~5, $\Delta_1 \approx 0.7\Delta_{\rm max}$.
The dependence of $\Delta_1$ on the concentration is given in Fig.~8.
Comparing the plots of the gap $\Delta_1$ and the chemical
  potential $\mu$ (Fig.~7), we conclude that $\Delta_1 \sim 0.7\mu$.
This is really a very strong coupling limit and
 virtually all holes are involved in pairing.
This is to be contrasted with the usual situation when only
 a small portion of electrons $\frac{\omega_D}{\epsilon_F}$ take part in
 pairing and the gap is proportional to the Debye frequency.

The $g$-wave pairing is weaker and we will not present complete results
 for this case.
Due to the above mentioned similarity of the long-range (small $q$)
 behavior of the $d$ and $g$ waves which arises from having the same number
 of zeros at the Fermi surface, the value of the gap for the $g$ wave
 for the above parameters is of the same order as for the $d$ wave.
It is also interesting that the $g$ wave does not depend on details
 of the contact part of the interaction (\ref{Vkk})
 while the $d$-wave gap is substantially suppressed by adding repulsion
 to the short-range interaction.
Thus, under certain conditions the $g$-wave solution may
 be relevant to the problem.
Table I gives more information about solutions at several parameters
 including the difference of the free energy
 $F = \langle \Psi | H - \mu N_h | \Psi \rangle$
 ($N_h$ is the number of holes)
 between the superconducting and normal states,
\begin{equation} \label{F}
F_S - F_N =  2 \sum_{\bf k}
          \xi_{\bf k} v^2_{\bf k}
         + {1\over 2} \sum_{\bf k, k'} V_{\bf k, k'}
         u_{\bf k} v_{\bf k} u_{\bf k'} v_{\bf k'}.
\end{equation}
It is convenient to calculate free energy
  per {\em hole} and use the difference
 $f_S - f_N = {1 \over {N\delta} } (F_S - F_N)$.

\section{Critical temperature}
\noindent
Due to the condition (\ref{omega}) the spin-wave frequency in the
 hole-hole interaction (\ref{Vkk}) is large in comparison with the hole
 excitation energy: $\omega_{\bf q} \gg E_{\bf k}$.
It means that retardation is small and
  the interaction is almost instantaneous.
It is well known that in this case the equation for the
 gap at $T \neq 0$ is
\begin{equation}\label{DeltaT}
  \Delta_{\bf k}= -\frac{1}{2} \sum_{\bf k'} V_{\bf kk'}
  \frac{\Delta_{\bf k'}}{E_{\bf k'}} \tanh {{E_{\bf k'}}\over{2T}}.
\end{equation}
In Fig.~9 we present the calculated dependence of $\Delta_1$
 on temperature at hole concentration $\delta=0.1$.
Figure~10 gives the dependence of the critical temperature $T_c$
 on hole concentration.
The approximate relation~(\ref{dtc}) derived analytically
 in Ref.~\onlinecite{Fla4} is qualitatively fulfilled.
In real units ($J=0.15$ eV), Fig.~10 gives
 (taking $a=7$)
  $T_c = 51$ K at $\delta=0.1$ and $T_c = 86$ K at $\delta=0.3$.
Let us stress that in our calculation we do not use any fit.
The only input is the values of $t$ and $J$.

\section{Conclusions}
\noindent
Using the single spin-wave exchange mechanism suggested in
 Refs.~\onlinecite{Kuch3,Fla4} we carried out a numerical {\em ab initio}
 calculation of superconducting pairing in the $t$-$J$ model.
Both the magnitude of critical temperature and its dependence on hole
 concentration are in good agreement with experimental data.
The calculated critical temperature is still smaller
 than the highest critical temperature obtained in experiment.
However, this may be explained by not knowing the exact parameters.
By a relatively small variation of parameters we can get
 $T_c =$ 100--150 K.

The most important remaining problem
  is the destruction of long-range antiferromagnetic order.
Following experimental results\cite{Birg} we have assumed
 that antiferromagnetic order is preserved at distances
 $r \lesssim 1/p_F$.
The behavior at larger distances is an open question in the present paper.

\acknowledgments

\noindent
We are very grateful to V. V. Flambaum, M. Yu.\ Kuchiev, V. F. Dmitriev,
 M. Mostovoy, and V. B. Telitsin for valuable discussions.
This work was supported in part by
 the Council on Superconductivity of the Russian Academy of Sciences,
    Grant No.~93197,
 Russian Foundation for Fundamental Research,  Grant No.~94-02-03235,
 Competition Center for Natural Sciences at St.~Petersburg State
    University, Grant No.~94-5.1-1060; and
 a grant from the Australian Research Council.

\tighten

\begin{table}
\caption{Influence of short-range interaction.
Changes in the short-range interaction are introduced by increasing the
 parameter $\alpha$ [see discussion after Eq.~(\protect\ref{14})]
 which makes the contact interaction more repulsive.
All data are for
 $T=0$, $a=7$, $\mu=0.058$ (this is the chemical potential at which
 in the absence of interaction the hole concentration
 would be $\delta_0=0.05$).}

\begin{tabular}{l l l l l l}
\multicolumn{1}{c}{Symmetry} &
\multicolumn{1}{c}{$\alpha$} &
\multicolumn{1}{c}{$f_S-f_N$} &
\multicolumn{1}{c}{$\delta$} &
\multicolumn{1}{c}{$\Delta_{\rm max}$} &
\multicolumn{1}{c}{$\Delta_1$} \\ \hline

$d$-wave& $\alpha=0.6$ & $-4.42\times 10^{-3}$ & 0.0606 & 0.0523 & 0.0377\\
$d$-wave& $\alpha=0.8$ & $-2.01\times 10^{-3}$ & 0.0552 & 0.0317 & 0.0222\\
$g$-wave& $\alpha=0.6$ & $-1.79\times 10^{-3}$ & 0.0547 & 0.0259 & 0.0210\\
$g$-wave& $\alpha=0.8$ & $-1.79\times 10^{-3}$ & 0.0547 & 0.0259 & 0.0210\\
\end{tabular} \end{table}


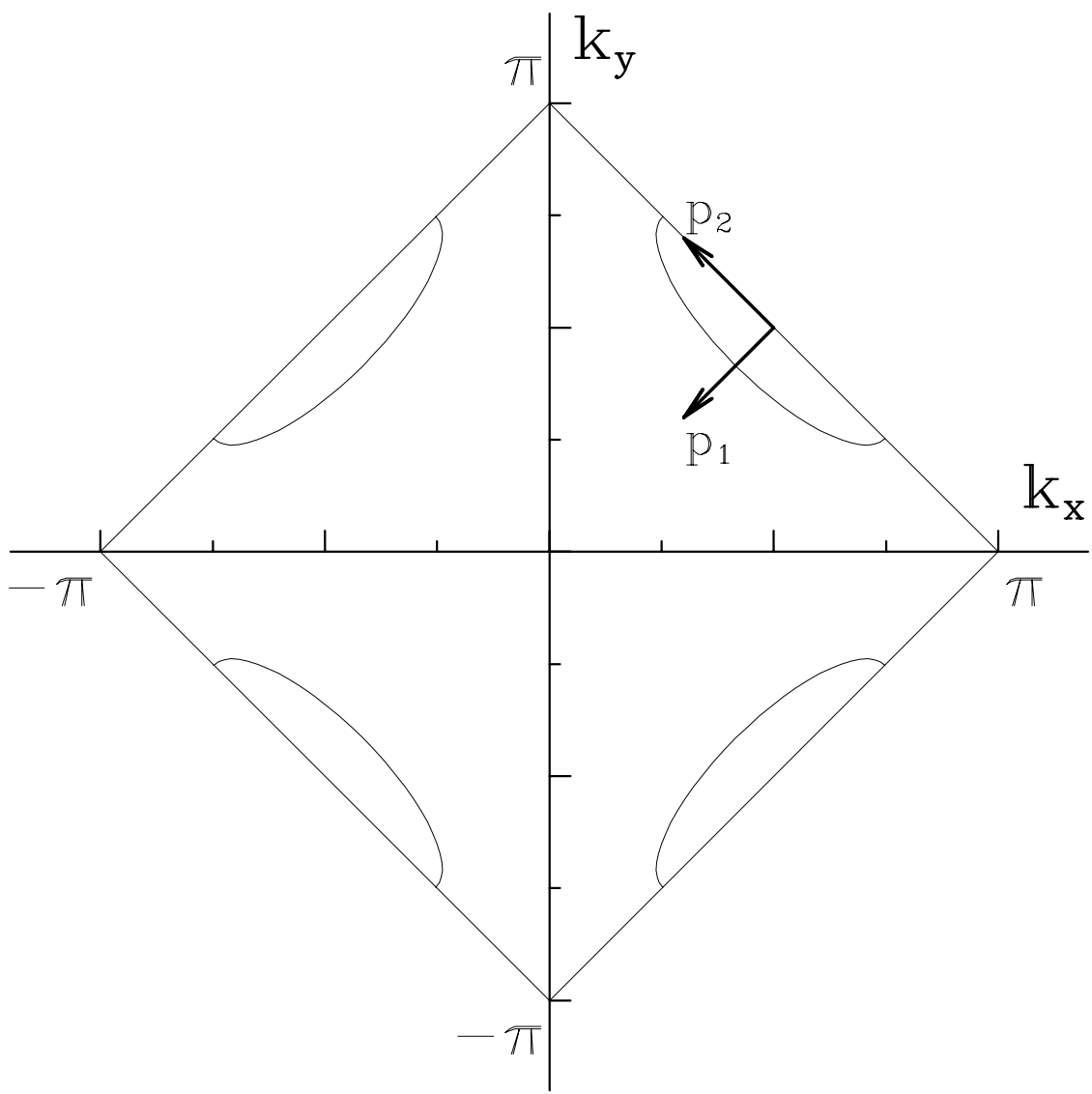
FIG. 1. The Brillouin zone of a hole in the $t$-$J$ model.

FIG. 2. The plot of the hole--spin-wave coupling constant $f$.

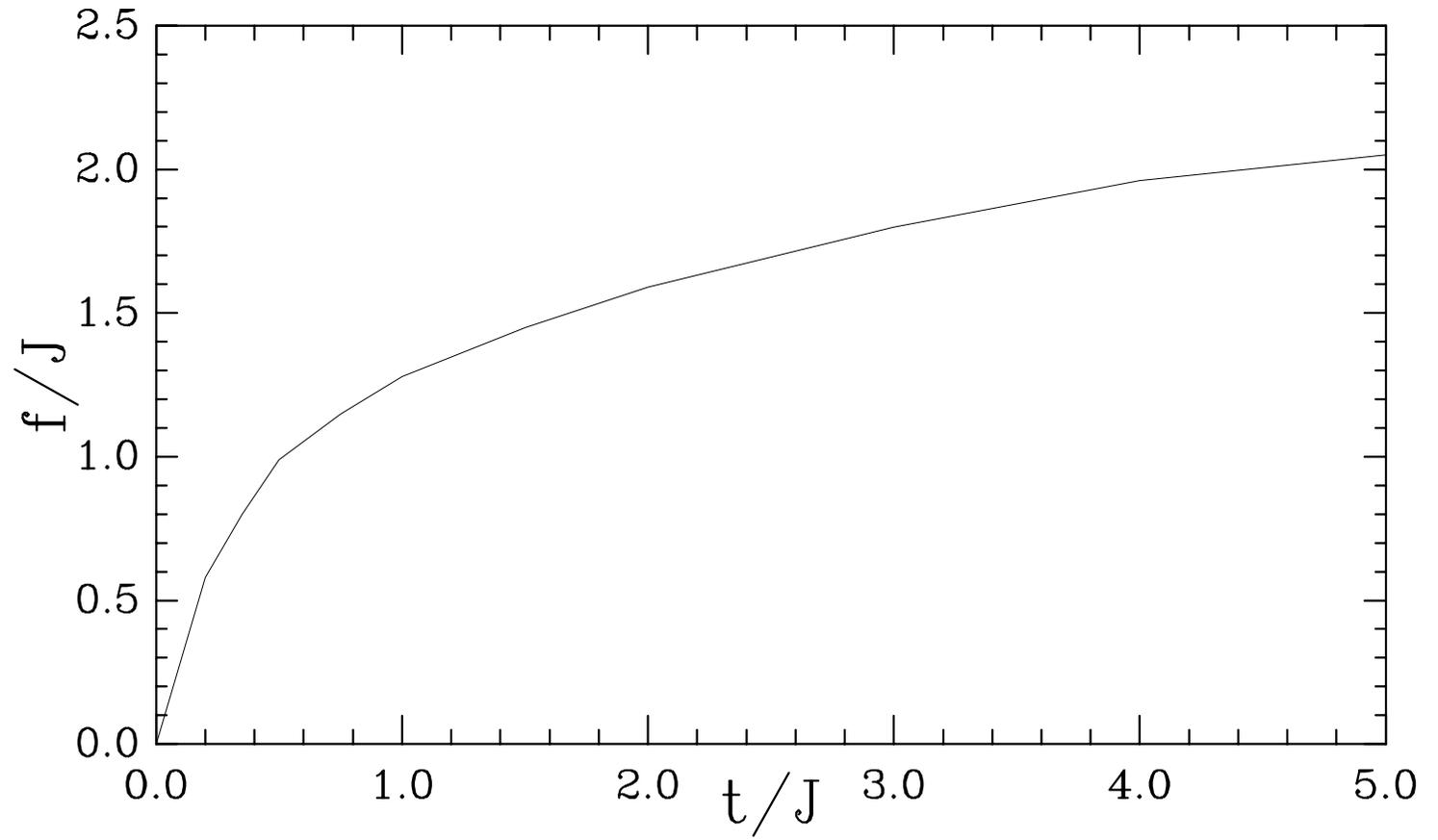
FIG. 3. Interaction between two holes via a single spin-wave exchange.

FIG. 4. The gap symmetry.
  (a) $B_1$ type ($d$ wave),
  (b) $A_2$ type ($g$ wave).

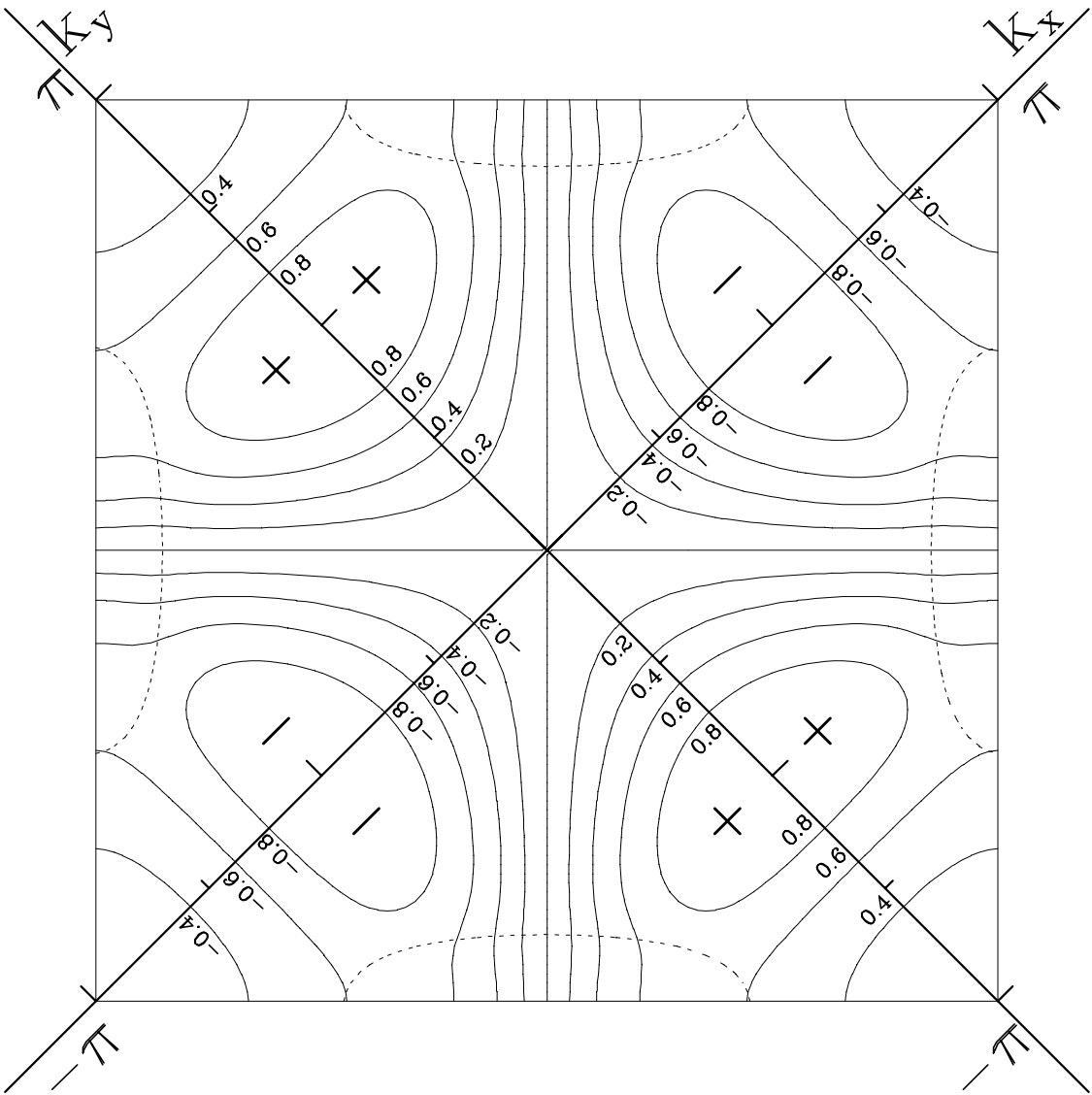
FIG. 5. The contour plot of the $d$-wave gap for $t/J=3$, the mass ratio
 $a=\beta_1/\beta_2=7,$ and hole concentration $\delta=0.1$.
The levels are presented in units of the gap maximal value which at
 these parameters is equal $\Delta_{\rm max}=0.0661$.
 Dashed curves represent the
Fermi surface for the case when one considers the holes like an ideal gas.

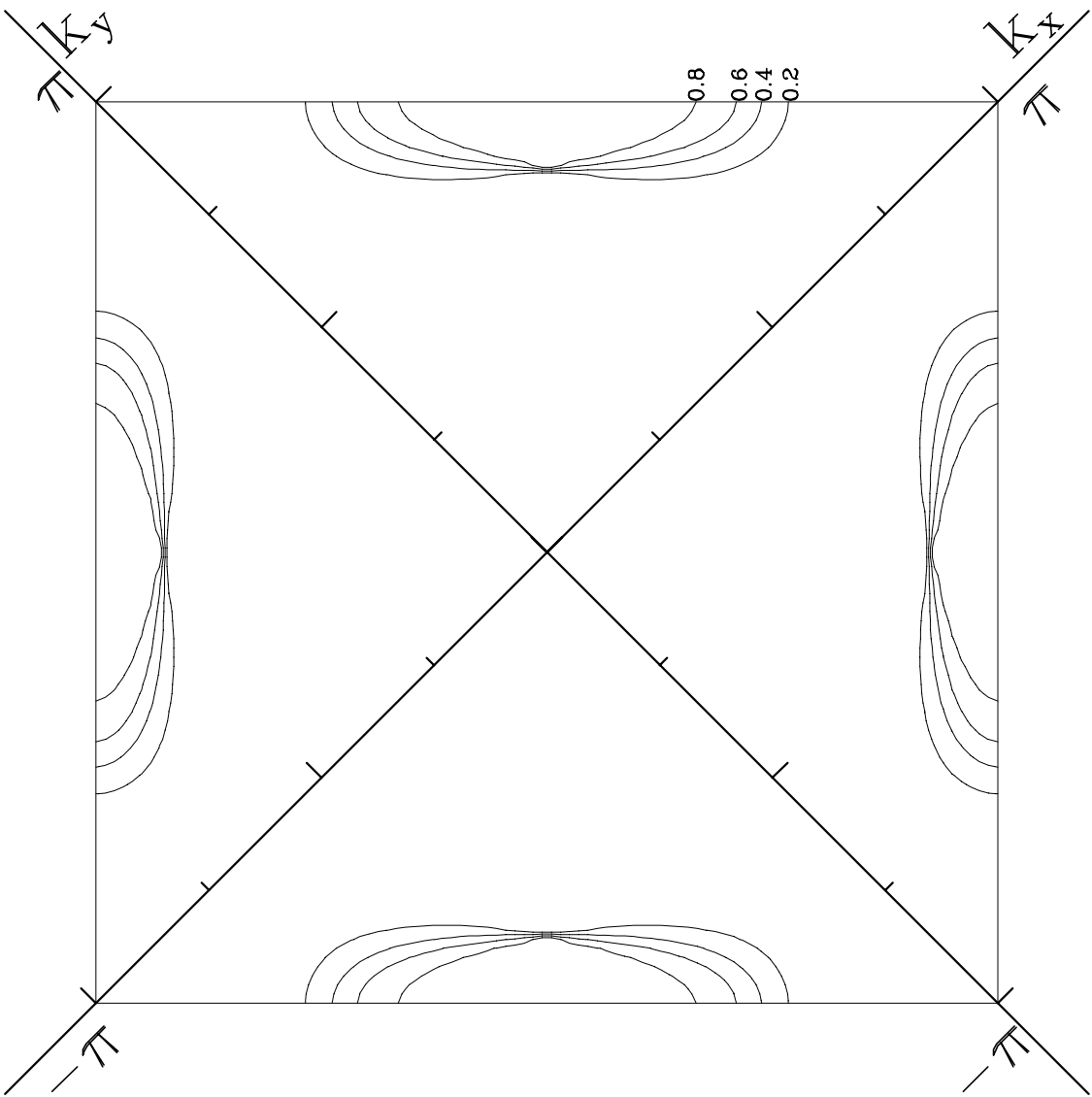
FIG. 6. The contour plot of a single hole quantum state mean occupation number
 $v_{\bf k}^2$. Parameters are the same as in Fig. 5.

FIG. 7. The chemical potential $\mu$ as a function of hole
 concentration $\delta$. Dashed curves correspond to an ideal gas of holes.
Deviation of dashed curves from linear dependence (\ref{ep})
 is due to the deviation of
 the dispersion relation (\ref{6}) from the quadratic expansion (\ref{7}).
The dependence with pairing taken into account is presented by solid lines.
All the curves correspond to $t/J=3$.
The mass ratio is (from top to bottom) $a=\beta_1/\beta_2=5,7,9$.

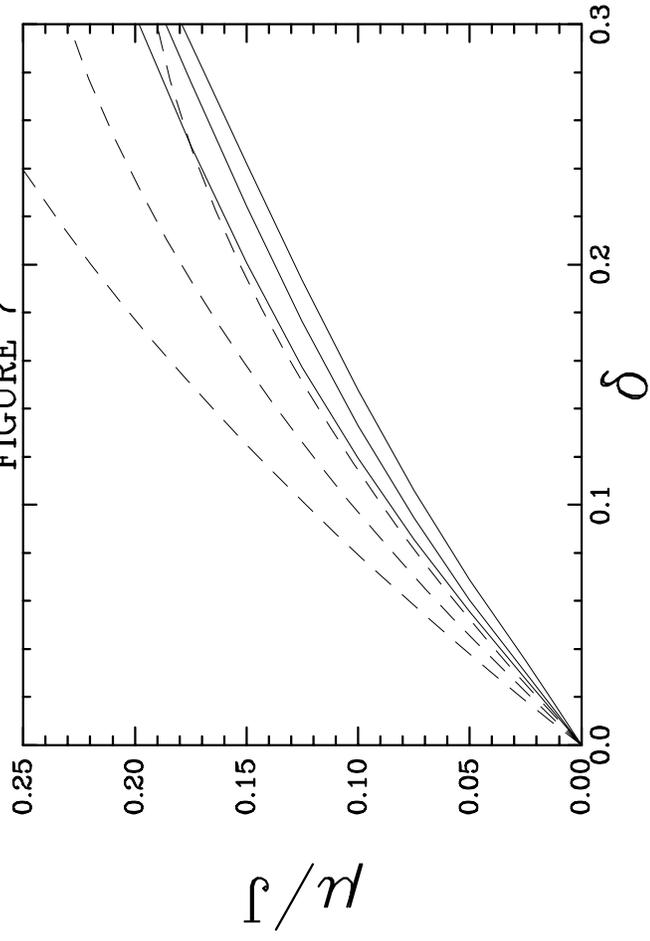
FIG. 8. The maximal value of the gap on the Fermi surface $\Delta_1$
 vs hole concentration for $t/J=3$;
 the mass ratio is (from top to bottom) $a=\beta_1/\beta_2=5,7,9$.

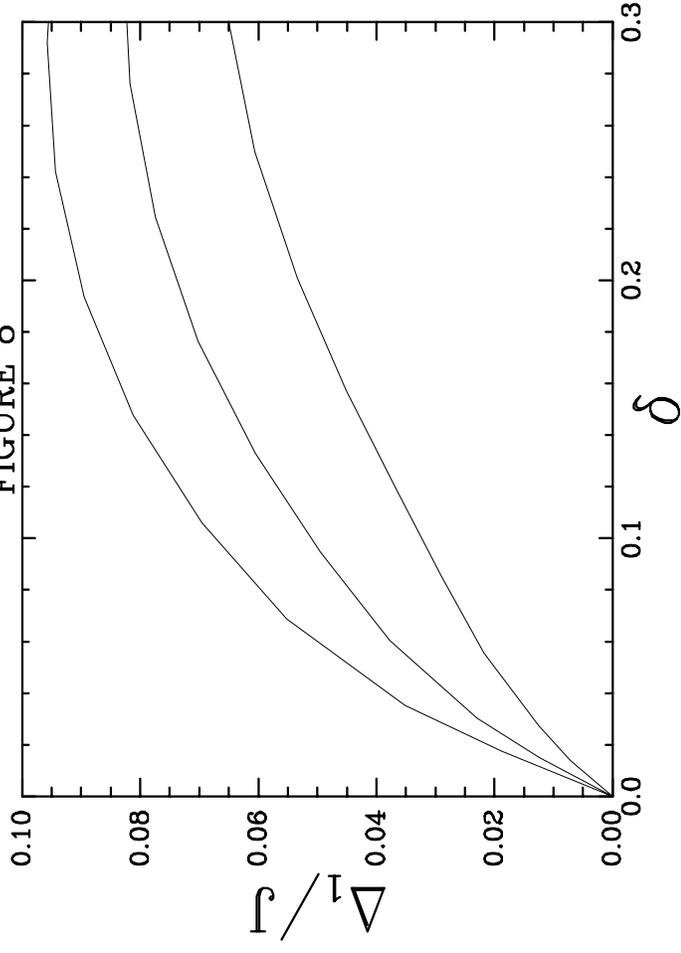
FIG. 9. The maximal value of the gap on the Fermi surface $\Delta_1$
 as a function of temperature. The hole concentration
 is $\delta=0.05$, $t/J=3$, the mass ratio $a=\beta_1/\beta_2=7$.

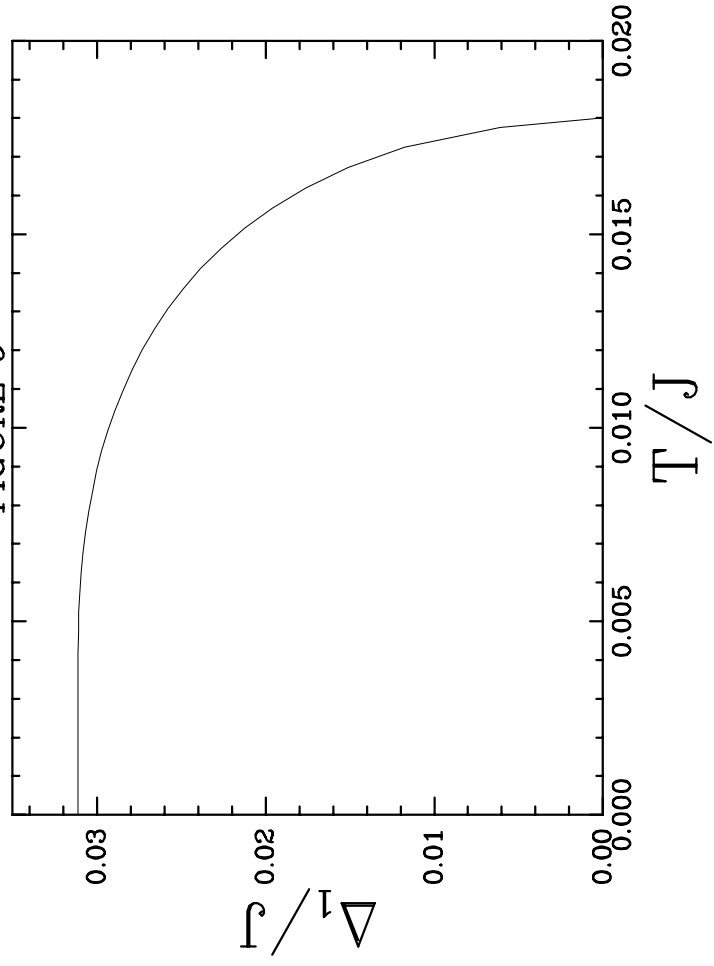
FIG. 10. The critical temperature vs hole concentration. $t/J=3$,
 the mass ratio $a=\beta_1/\beta_2 = 7$.


\begin{thebibliography}{99}
\bibitem{Sch9} J. R. Schrieffer, X. C. Wen, and S. C. Zhang,
  Phys.\ Rev.\ B {\bf 39}, 11663 (1989).
\bibitem{Bic9} N. E. Bickers, D. J. Scalapino, and S. R. White,
  Phys.\ Rev.\ Lett. {\bf 62}, 961 (1989).
\bibitem{Kam1} A. Kampf and J. R. Schrieffer,
  Phys.\ Rev.\ B {\bf 41}, 6399 (1990).
\bibitem{Kam2} A. Kampf and J. R .Schrieffer,
  Phys.\ Rev.\ B {\bf 42}, 7967 (1990).
\bibitem{Bul1} N. Bulut and D. J. Scalapino,
  Phys.\ Rev.\ B {\bf 45}, 2371 (1992).
\bibitem{Bul2} N. Bulut, D. J. Scalapino, and R. T. Scalettar,
  Phys.\ Rev.\ B {\bf 45}, 5577 (1992).
\bibitem{Mon1} P. Monthoux, A. V. Balatsky, and D. Pines,
  Phys.\ Rev.\ Lett. {\bf 67}, 3448 (1991);
  Phys.\ Rev.\ B {\bf 46}, 14803 (1992).
\bibitem{Mon2} P. Monthoux and D. Pines,
  Phys.\ Rev.\ Lett. {\bf 69}, 961 (1992);
  Phys.\ Rev.\ B {\bf 47}, 6097 (1993);
  Phys.\ Rev.\ B {\bf 49}, 4277 (1994).
\bibitem{Dah3} T. Dahm, J. Erdmenger, K. Scharnberg, and C. T. Rieck,
  Phys.\ Rev.\ B {\bf 48}, 3896 (1993).
\bibitem{Kuch3} M. Yu. Kuchiev and O. P. Sushkov,
   Physica C {\bf 218}, 197 (1993).
\bibitem{Fla4} V. V. Flambaum, M. Yu. Kuchiev, and O. P. Sushkov,
  Physica C {\bf 227}, 267 (1994).
\bibitem{Dag5} E. Dagotto, A. Nazarenko, and A. Moreo,
   Phys. Rev. Lett., to be published.
\bibitem{Birg} R. J. Birgeneau and G. Shirane, in
  {\em Physical Properties of High Temperature Superconductors},
  edited by D. M. Ginsberg (World Scientific, Singapore, 1989).
\bibitem{Manousakis} E. Manousakis, Rev.\ Mod.\ Phys. {\bf 63}, 1 (1991).
\bibitem{Dag4} E. Dagotto,
  Rev.\ Mod.\ Phys. {\bf 66}, 763 (1994) and references therein.
\bibitem{Kane} C. L. Kane, P. A. Lee, and N. Read, Phys. Rev. B
  {\bf 39}, 6880 (1989).
\bibitem{Dag0} E. Dagotto, R. Joynt, S. Bacci, and E. Cagliano,
  Phys.\ Rev.\ B {\bf 41}, 9049 (1990).
\bibitem{Mart1} G. Martinez and P. Horsch, Phys.\ Rev.\ B {\bf 44}, 317 (1991).
\bibitem{Liu2} Z. Liu and E. Manousakis, Phys.\ Rev.\ B {\bf 45}, 2425 (1992).
\bibitem{Sus1} O. P. Sushkov, Phys.\ Lett.\ A {\bf 162}, 199 (1992).
\bibitem{Sus2} O. P. Sushkov, Solid State Commun. {\bf 83}, 303 (1992).
\bibitem{She2} A. V. Sherman, Phys.\ Rev.\ B {\bf 46}, 6400 (1992).
\bibitem{Giam3} T. Giamarchi and C. Lhuillier,
   Phys.\ Rev.\ B {\bf 47}, 2775 (1993).
\bibitem{Suhf} O. P. Sushkov, Phys.\ Rev.\ B {\bf 49}, 1250 (1994).
\bibitem{Suh3} O. P. Sushkov, Physica C {\bf 206}, 264 (1993).
\bibitem{Cher3} A. L. Chernyshev, A. V. Dotsenko, and O. P. Sushkov,
  Phys.\ Rev.\ B {\bf 49}, 6197 (1994).
\bibitem{Ede2} R. Eder, Phys.\ Rev.\ B {\bf 45}, 319 (1992).
\bibitem{Bon2} M. Boninsegni and E. Manousakis,
  Phys.\ Rev.\ B {\bf 47},  11897 (1993).
\bibitem{Poilblanc} D. Poilblanc, Phys.\ Rev.\ B {\bf 48}, 3368 (1993).
\bibitem{PRD3} D. Poilblanc, J. Riera, and E. Dagotto,
  Phys.\ Rev.\ B {\bf 49}, 12318 (1994).
\bibitem{Sher} A. V. Sherman, Physica C {\bf 211}, 329 (1993).
\bibitem{Esk0} H. Eskes, G. A. Sawatzky, and L. F. Feiner,
  Physica C {\bf 160}, 424 (1989).
\bibitem{Fla1} V. V. Flambaum and O. P. Sushkov,
  Physica C {\bf 175}, 347 (1991).
\bibitem{BCh3} V. I. Belinicher and A. L. Chernyshev,
  Phys.\ Rev.\ B {\bf 47}, 390 (1993);
  Phys.\ Rev.\ B {\bf 49}, 9746 (1994);
    V. I. Belinicher, A. L. Chernyshev, and L. V. Popovich,
  Phys.\ Rev.\ B {\bf 50}, 13~768 (1994).
\end{thebibliography}
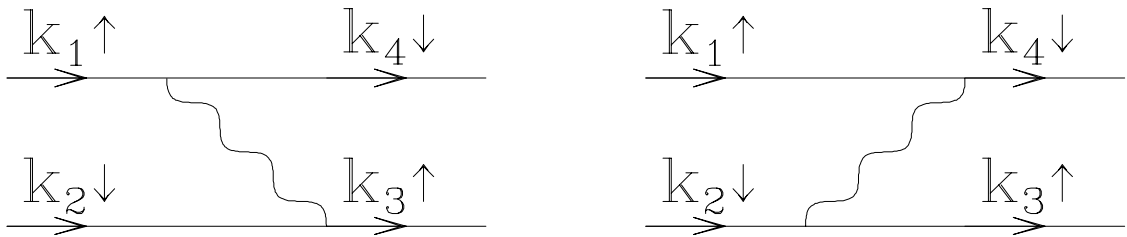
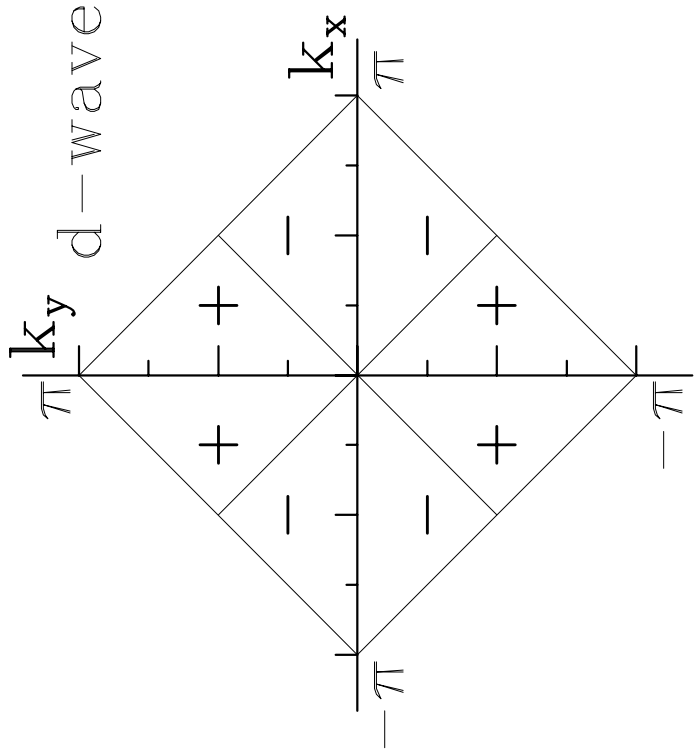
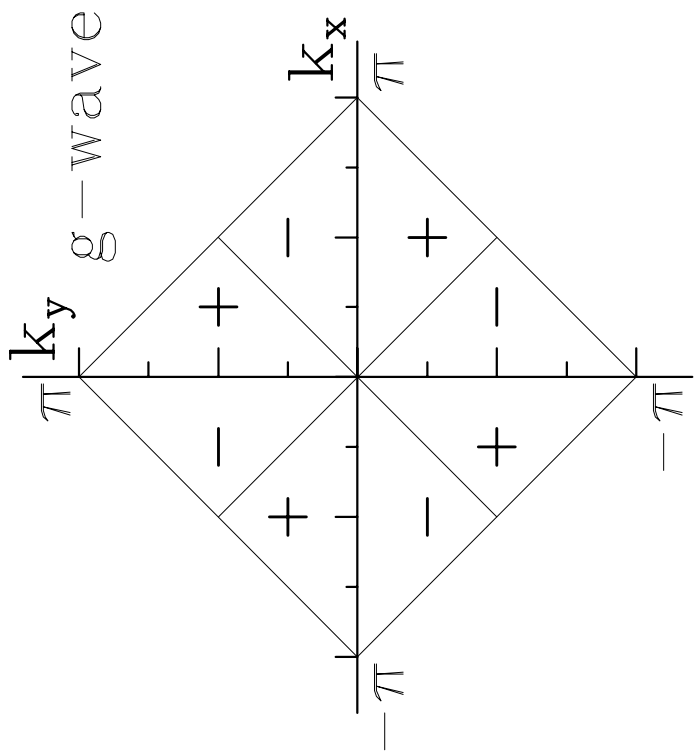
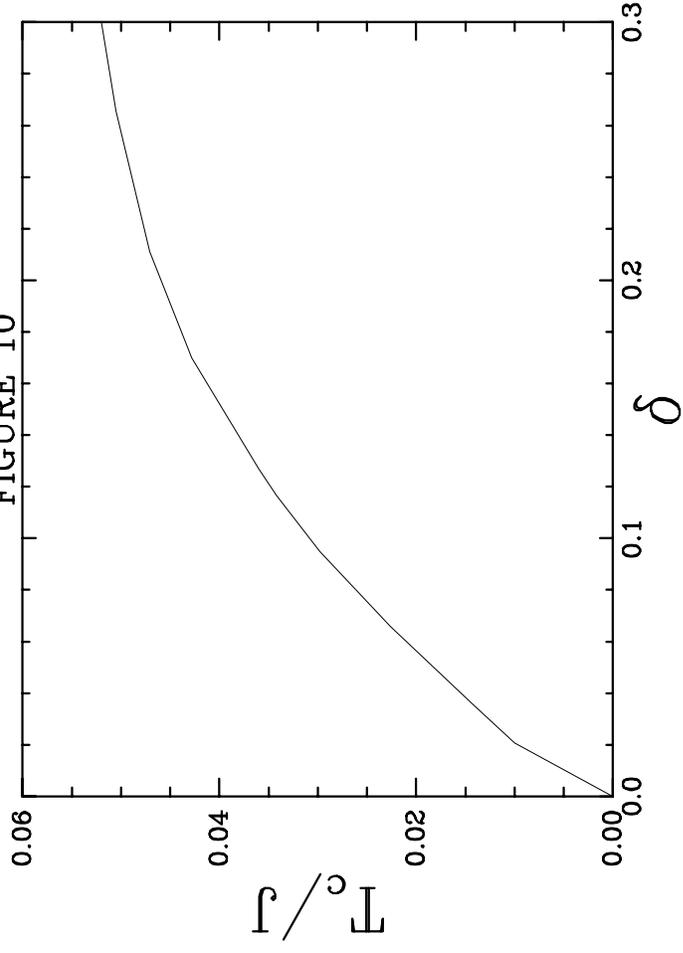
\end{document}